**Modification of poly(*L*-lactide) and polycaprolactone bioresorbable polymeric materials by RF plasma discharge: a preliminary study evaluating EA-hy 926 cell attachment**

E.N. Bolbasov[1], M. Rybachuk[2], A.S. Golovkin[3], L.V. Antonova[3], E.V. Shesterikov[1], A.I. Malchikhina[1], V.A. Novikov[4], Y.G. Anissimov[5] and S.I. Tverdokhlebov[1a*]

[1]*Tomsk Polytechnic University, 30 Lenin Avenue, Tomsk 634050, Russian Federation*

[2]*Griffith University, School of Engineering, Engineering Dr., Southport, QLD 4222, Australia*

[3]*Kuzbass Cardiology Centre - Research Institute for Cardiovascular Diseases, Russian Academy of Medical Sciences (Siberian Branch), 6 Sosnovy Blvd, Kemerovo 650002, Russian Federation*

[4]*Tomsk State University, 36 Lenin Avenue, Tomsk 634050, Russian Federation*

[5]*Griffith University, School of Biomolecular and Physical Sciences, Engineering Dr., Southport, QLD 4222, Australia*

[*]*communicating author: tverd@tpu.ru*

ABSTRACT

Surface modification of poly(*L*-lactide) (*L*-PLA) and polycaprolactone (PCL) bioresorbable polymers by radio-frequency thermal glow discharge plasma is reported. Improved biocompatibility of *L*-PLA and PCL materials was obtained by employing hydroxyapatite target sputtering in $Ar^+$ plasma as evidenced by the change of *L*-PLA and PCL properties from highly hydrophobic to hydrophilic, with absolute wettability obtained for both materials, and enhanced endothelial hybrid cell line EA-hy 926 attachment to modified surfaces. For the latter, surface properties that suppress adverse cellular responses (*e.g.* apoptosis, necrosis) were also attained. Surface roughness and surface free energy were found to increase significantly for both polymers under prolonged plasma exposure and, a longer chain aliphatic PCL were found to display a marginally better post plasma-treated biocompatibility compared to *L*-PLA.

MAIN TEXT

Applications of biodegradable materials such as poly(*L*-lactide) $(C_3H_4O_2)_n$ aliphatic polyester (*L*-PLA) and polycaprolactone $(C_6H_{10}O_2)_n$ (PCL) in an environment where attachment and proliferation of living cells is important necessitate specific surface properties, namely, the surface roughness, to be within tens of nm (normally in 20 – 80 nm range and tailored to a specific cell line) and to be highly hydrophilic.[1,2] Both surface topology and physio-chemical properties determine of how cells are attached to the surface defining cell polarisation, cell morphology, spatial orientation of cytoskeleton components, the degree of order of intracellular transport and many other important parameters.[3] Recent works on surface treatment of *L*-PLA and PCL polymers by ionised media revealed that it generally leads to an increase of surface roughness and reduces surface free energy, which, consequently, results in an improved attachment of various biological cells on the surface.[3,4] However, despite a range of plasma treatment methods (*i.e.* thermal and non-thermal plasma corona discharge, dielectric barrier discharge, etc.) available inherent hydrophobicity and high surface-free energy properties of *L*-PLA and PCL were found to be poorly amendable in a controlled manner and still present a major drawback for many application[5-8]. While non-thermal plasma surface treatments are preferred for simplicity, modification of *L*-PLA and PCL under thermal plasma conditions are less popular owing to inherent difficulties associated with identifying appropriate plasma conditions and defining the complimenting target material for a specific (bio)-material surface treatment.

This Letter is aimed to provide initial results on the study of the effects of treatment of *L*-PLA and PCL polymers employing a radio-frequency (RF) glow discharge plasma operated under $Ar^+$ target sputtering mode in order to reduce the degree of hydrophobicity and surface-free energy to ultimately improve living cell affinity. Hydroxyapatite (HA) $Ca_5(PO_4)_3(OH)$ was chosen as a sacrificial target owing to its natural (bio)-compatibility and ability to facilitate biological functions of various implant materials by plasma sputtering processes [9-11]. HA being a dielectric serves as an ideal target material for an RF excited, plasma-assisted, sputtering of *L*-PLA and PCL to form a (bio-)ceramic/(bio-)resorbable polymer interface in support of live cell attachment studies.

PURASORB® PL65 (Corbion Purac) with inherent viscosity (IV) at midpoint $IV_{L-PLA}$ = 6.5 dL/g, and Policaprolactone PURASORB® PC12 $IV_{PCL}$ = 1.2 dL/g (Corbion Purac) were used as precursor stock polymers to prepare a 4% solution of *L*-PLA and PCL polymers, respectively, in dichloromethane $CH_2Cl_2$ (DCM) (Panreac Química S.L.U.). A 4% polymer solution of 12 (±1) g was casted at room temperature (RT) into a glass vial to solidify. After 24 hours when most of DCM was vaporized, the vial was filled with a distilled milli-Q water to separate the formed polymer film from the vials' inner surface, followed by air-drying at 35° C for 24 hours to remove residual water and DCM. As-prepared *L*-PLA and PCL polymer films were subjected to an RF glow discharge plasma treatment in an RF system described earlier by Tverdokhlebov *et al.*[12]. During the operation the specific RF power was set to 5 W/cm$^2$ to maintain electron density within ~ $10^9 - 10^{10}$ ppcm$^3$ range and target-to-substrate distance was extended to 16 cm to facilitate enhanced coagulation of elemental plasma and target species under collisional plasma conditions.[13] Film exposure to plasma was 30, 60 and 150 s. Topology of plasma treated surfaces was studied employing atomic force microscopy (AFM) (Solverb - HV, NT-MDT) in air at RT using tapping mode; Gwyddion 2.25[14] analytical package was used to estimate root mean square values of surface roughness, *RMS*. Wetting tests were performed with propan-1,2,3-triol (glycerol) and Milli-Q water using a contact angle goniometer (Drop Shape Analyzer - DSA25, KRÜSS GmbH) whereas the surface free energy, $\gamma$, was estimated using standard Owens, Wendt, Rabel and Kaelble (OWRK) method[15]. Endothelial hybrid cell line EA-hy 926 provided by Cora-Jean S. Edgell, University of North Carolina, USA were used in cell affinity studies. For the latter, 15 mm diameter flat discs (1.8 cm$^2$) were stencilled from as-casted polymer films and placed in a 24-well plate. The cells were cultivated in a 1:1 mixture of Dulbecco's Modified Eagle's Medium (DME) and Ham's F-12 Nutrient Mixture (DME F12) (Sigma-Aldrich), with an addition of 1% HEPES buffer, 10% fetal bovine serum, 1% *L*-glutamine, 100 ppm penicillin, 0.1 mg/mL streptomycin, 0.1 mg/mL amphotericin B, and a hypoxanthine, minopterin, thymidine mixture (HAT) (Sigma-Aldrich) in an environment containing 5% $CO_2$ at 37º C for 72 hours.

The obtained EA-hy 926 cells were labelled using fluorescent cell linker dye PKH26 (Sigma-Aldrich) and bis-Benzimide H 33342 trihydrochloride nucleic acid stain (Sigma-Aldrich). Estimation of cell population was carried out employing an inverted Axio Observer Z1 (Carl Zeiss) microscope; cell population density was averaged at 1 mm$^2$ resolution with the median value, $ME$, reported together with the 1$^{st}$ (25%) and 3$^{rd}$ quartile (75%) quartiles. FACSCalibur$^{TM}$ flow cytometer (BD Biosciences) was used for cell viability studies; for the latter the cells were removed from the $L$-PLA and PCL materials using a 0.5% Trypsin – EDTA solution (Sigma-Aldrich), centrifuged under 716 g for 10 min and re-suspended in a 1 ml of original culture medium; this step was followed by staining the cells with Annexin A5, phycoerythrin (PE) and 7-AAD (7-amino-actinomycin D) (BD Biosciences). Control group of EA-hy 926 cells was cultivated on stock (untreated) $L$-PLA and PCL substrates.

Figures 1 and 2 show the results of surface topology studies for stock $L$-PLA and PCL samples (see FIG. 1a for $L$-PLA, FIG. 2a for PCL), and for plasma-modified samples; the first column displaying AFM topographic images, the second showing a 2D fluorescent map of labelled EA-hy 926 cells; a vertical profile corresponding to wettability images $L$-PLA (FIG. 1) and PCL (FIG. 2) with H$_2$O and glycerine droplets displayed on the right. A summary of topological features for stock and plasma-treated $L$-PLA and PCL is given in Table I, where changes for $R_a$, contact angles of water, $\theta_{H2O}$, and glycerol, $\theta_{gl}$, surface free energy, $\gamma$, including apolar, $\gamma^{LW}$, component accounting for Liftshitz/van der Waals (LW) type interactions and polar, $\gamma^{AB}$, component accounting for acid-base (AB) or donor-acceptor type interactions are displayed as a function of the exposure time, $t$. The obtained results indicate that the measured values for surface roughness gradually increase to a maximum of ~30 nm for $L$-PLA and ~70 nm for PCL with plasma exposure due to thermo-chemical degradation of the polymeric material mostly caused by increased presence of free H- bonds and plasma and target species bombardment. The latter are reactive, positively charged $CaO^+$, $Ca^{2+}$, $PHO^+$, $PO^+$, $P^+$ species sputtered out from the HA target which are attracted by the negatively charged surface of $L$-PLA and PCL. The result of these interactions is the formation of an amorphous calcium aliphatic compound with significantly altered populations of carbonyl (-C=O), hydroxyl (-OH) functional groups and, hydroxyapatite pre-nucleation layer on *L*-PLA and PCL materials. Earlier, Ágata de Sena *et al.*[16] confirmed the formation of the amorphous phase developing at early film growth stages for thermal HA sputtering on metal substrates (*e.g.* Ti) as $CaTiO_3$. We postulate that these processes occur simultaneously resulting in the formation of an amorphous (bio-)ceramic/(bio-)resorbable polymer interface such as *a*-Ca-(*L*-PLA)-$O_3$ on *L*-PLA, and *a*-Ca-(PCL)-$O_3$ on PCL material. Plasma treatment of both *L*-PLA and PCL materials gradually change their surface properties from hydrophobic, with surface energy values of ~20 mJ/m$^2$ for *L*-PLA and ~30 mJ/m$^2$ for PCL to largely hydrophilic, with a maximum values of $\gamma$ of ~73 mJ/m$^2$ reported for both materials indicating absolute wettability. Since it is established that surface free energy of hydrocarbons $C_nH_{2n-2}$ is strongly related to C- saturation[17], the increase of hydrophilic properties in *L*-PLA and PCL materials can be attributed to strong de-hydrogenation process occurring during the plasma treatment. Moreover, since amorphous solids traditionally exhibit lower surface energy values than crystalline counterparts, we postulate that *L*-PLA and PCL surface properties undergo marked transition from an amorphous to a crystalline state, which further supports the probability of formation of a (bio-)ceramic/(bio-)resorbable polymer interface. The marked increase of $\gamma^{AB}$ (*i.e.* acid-base type interactions) values for both materials reveals that plasma treatment promotes high intermolecular force interactions on the surface, while having a minor, insignificant, effect on apolar surface properties of *L*-PLA and PCL materials. The latter is evidenced by minimal changes of $\gamma^{LW}$ component (*i.e.* Liftshitz/van der Waals type interactions). Two reasons give rise to high intermolecular forces: one is the presence of local reactive molecular domains participating in radical formation, molecular cross-linking and/or C= bond formations on polymer surface under plasma irradiation, the other are the interactions with the positive (and neutral) species from the HA target noted earlier. High intermolecular force interactions on plasma treated *L*-PLA and PCL can contribute towards the increase of C-O and O-C=O functional groups with additional O- ions supplied by the sacrificial HA target. We attribute a major increase of surface roughness in PCL relative to *L*-PLA to inherent differences in their molecular structure: *L*-PLA while lacking a longer C=C backbone is less predisposed to severe thermal and physio-chemical modifications than PCL counterpart, and as PCL exhibits lower solubility, it is naturally more prone to molecular crosslinking during processing. PCL therefore displays more uneven surface with the most pronounced (*i.e.* higher value) $R_a$ under plasma treatment.

Living cell culture experimental data is summarized in Table II. It shows an increased affinity of EA-hy 926 cell attachment to plasma modified surfaces and retrospectively indicates that the increase of surface roughness and modification of surface functional groups is beneficial to overall cell viability. No significant effect is noted for cell apoptosis and/or necrosis measurements revealing that plasma-induced changes of surface morphology and the number and distribution of functional groups on *L*-PLA and PCL convey no known cell toxicity factors. Notably, a longer chain aliphatic PCL polymer was found to display a marginally better post plasma-treated biocompatibility compared to shorter-chain *L*-PLA.

In summary, we have shown that surface properties of *L*-PLA and PCL bioresorbable polymers can be successfully modified employing thermal glow discharge plasma sputtering hydroxyapatite target. Surface roughness, surface free energy could be altered by plasma treatment to achieve maximum wettability of *L*-PLA and PCL materials enhancing endothelial hybrid cell line EA-hy 926 attachment. We postulate the formation of a (bio-)ceramic/(bio-)resorbable polymer interface as the result of plasma-surface interaction and suggest that the reported complimentary (bio-)target sputtering approach is not exclusive to *L*-PLA and PCL polymers, but can aid in modification of other bioresorbable materials.

This work was financially supported by the Russian Foundation for Basic Research (projects #13-08-98052 r_sibir_a; #13-08-90743). M.R. acknowledges Griffith for lectureship program and start-up funding.

Table I. Surface roughness, contact angle wettability and surface free energy values for stock (untreated) and samples treated in plasma for 30, 60 and 150 s.

| Plasma exposure time, $t$, s | Surface roughness, $RMS$, nm | Contact angle | | Surface free energy | | |
|---|---|---|---|---|---|---|
| | | $\theta_{H2O}$, deg | $\theta_{gl}$, deg | $\gamma$, mJ/m$^2$ | $\gamma^{LW}$, mJ/m$^2$ | $\gamma^{AB}$, mJ/m$^2$ |
| poly(*L*-lactide) (*L*-PLA) | | | | | | |
| untreated | ≤ 5 | 88 ± 7 | 84 ± 6 | 20 ± 5 | 8 ± 2 | 12 ± 2 |
| 30 | ≤ 5 | 46 ± 3 | 45 ± 5 | 52 ± 3 | 13 ± 2 | 40 ± 2 |
| 60 | ≤ 5 | 34 ± 3 | 42 ± 2 | 65 ± 2 | 7 ± 1 | 58 ± 1 |
| 150 | 30 ± 5 | 24 ± 4 | 38 ± 2 | ≤ 73 | 6 ± 1 | 67 ± 2 |
| polycaprolactone (PCL) | | | | | | |
| untreated | 40 ± 5 | 77 ± 5 | 85 ± 3 | 33 ± 2 | ≤ 5 | 32 ± 2 |
| 30 | 45 ± 5 | 42 ± 3 | 47 ± 3 | 56 ± 4 | 12 ± 3 | 45 ± 2 |
| 60 | 55 ± 5 | 35 ± 4 | 40 ± 6 | 63 ± 4 | 9 ± 2 | 53 ± 2 |
| 150 | 70 ± 5 | 13 ± 3 | 25 ± 4 | ≤ 73 | 11 ± 2 | 62 ± 4 |

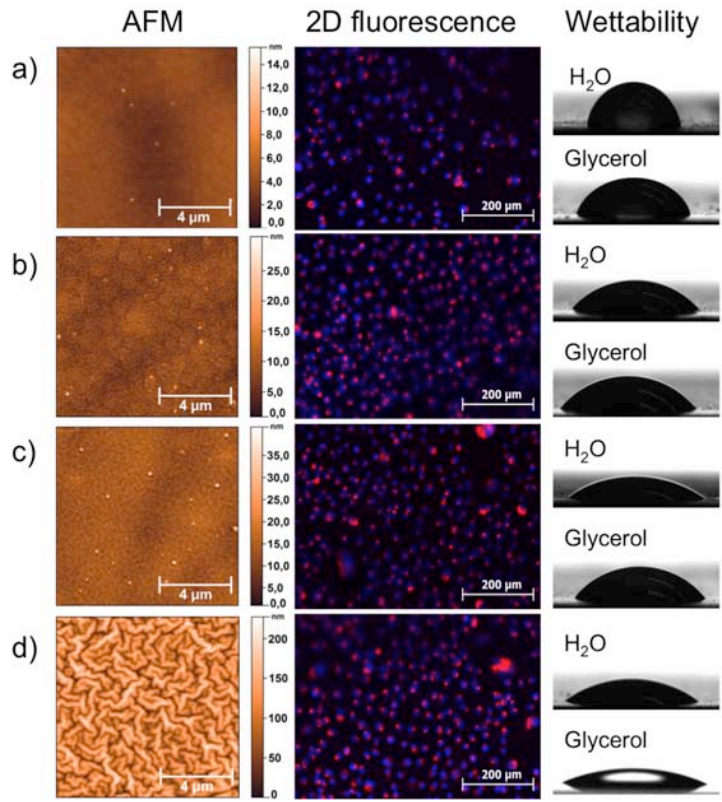

FIG 1. AFM topology, 2D fluorescence of labeled EA-hy 926 cells and wettability data for a) untreated *L*-PLA samples, and *L*-PLA samples treated in plasma for b) 30 sec, c) 60 sec and d) 150 sec

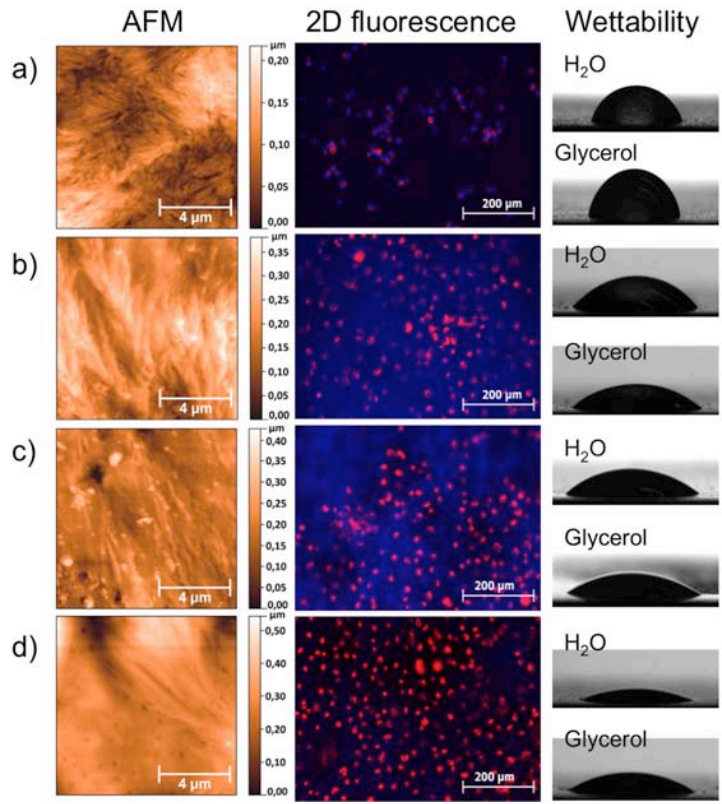

FIG 2. AFM topology, 2D fluorescence of labeled EA-hy 926 cells and wettability data for a) untreated PCL samples, and PCL samples treated in plasma for b) 30 sec, c) 60 sec and d) 150 sec

TABLE II. EA-hy 926 cell viability data for stock (untreated) and samples treated in plasma for 30, 60 and 150 s.

| Exposure time, $t$, s | Total EA-hy 926 population $ME$ (25%, 75%) | EA-hy 926 cell population (relative), $ME$ (25%, 75%)) | | | |
|---|---|---|---|---|---|
| | | Viable, % | Early apoptosis, % | Late apoptosis, % | Necrosis, % |
| poly($L$-lactide) ($L$-PLA) | | | | | |
| untreated | 409 (310; 471) | 91 (90; 91) | 7,4 (5,9; 8,9) | 0,7 (0,5; 0,8) | 1,0 (0,1; 1,9) |
| 30 | 775 (711; 823) | 93 (92; 94) | 5,3 (5,0; 5,6) | 0,8 (0,6; 1,0) | 0,7 (0,4; 1,0) |
| 60 | 741 (696; 828) | 93 (92; 94) | 5,1 (3,9; 6,4) | 0,8 (0,6; 1) | 0,7 (0,6; 0,8) |
| 150 | 624 (487; 738) | 92 (90; 92) | 7,5 (6,4; 8,6) | 0,5 (0,4; 0,6) | 0,3 (0,2; 0,5) |
| polycaprolactone (PCL) | | | | | |
| untreated | 242 (174; 300) | 90 (87; 91) | 9,3 (7,4; 11,3) | 0,6 (0,4; 0,8) | 0,4 (0,3; 0,5) |
| 30 | 417 (254; 694) | 84 (76; 91) | 14,9 (6,9; 22,9) | 0,6 (0,4; 0,9) | 0,8 (0,4; 1,1) |
| 60 | 501 (286; 669) | 90 (89; 91) | 7,8 (5,0; 9,3) | 0,8 (0,6; 1,0) | 1,5 (0,5; 2,3) |
| 150 | 715 (663; 760) | 92 (91; 92) | 6,5 (6,1; 6,8) | 0,8 (0,7; 0,9) | 0,6 (0,6; 0,6) |